\documentclass[prd,tightenlines,twocolumn]{revtex4}
\usepackage{amsmath}
\usepackage{amssymb}
\usepackage{graphicx}
\usepackage{bm}
\usepackage{enumerate}
\usepackage{color}
\newcommand{\be}{\begin{equation}}
\newcommand{\ee}{\end{equation}}
\newcommand{\bea}{\begin{eqnarray}}
\newcommand{\eea}{\end{eqnarray}}
\newcommand{\ba}{\begin{eqnarray}}
\newcommand{\ea}{\end{eqnarray}}

\begin{document}

\title{Comments on the SU(4) dark matter}
\author{ 
 Edward Shuryak }

\affiliation{Department of Physics and Astronomy, \\ Stony Brook University,\\
Stony Brook, NY 11794, USA}

%\date{\today}

\begin{abstract} 
We discuss possible scale of $SU(4)$ dark matter, in form of neutral baryons. We argue that
it is very likely that those would have time to cluster into large "nuclear drops" in which they are Bose-condensed.
 \end{abstract}

\maketitle
\section{Introduction}
The nature of dark matter remains subject to wide speculations, and experimental searches focus on mass 
and couplings in wide range of scales. The author, not being a specialist in the field, would not even attempt to
give references in respect to various theoretical ideas and current experimental limits.

These comments is focusing on one particular option, of a strongly interacting composite dark matter produced by hypothetical
QCD-like theories which, unlike QCD, have the lowest baryons neutral in electroweak charges. Furthermore, if the number of colors is even,
these baryons are likely to have spin zero, which reduces their interactions further.

One particular version of that, based on $SU(4)$ color group, has been studied rather extensively on the
lattice \cite{Appelquist:2014jch}, addressing masses of the mesons and baryons of this theory. 

We will consider the number of fermionic flavors be $N_f=2$ or higher. In this case 
there exist flavor-asymmetric spin-zero diquarks, which, like in QCD, most likely are deeply bound, and therefore
spin-zero baryon can be approximated by a pair of diquarks bound by the confining color flux tube.

\section{What may the natural scale of the SU(4) theory be?}
While this question is obviously most important, I have not seen its discussion
in literature. 

Perhaps the first idea one may try is to assume {\em Grand Unification of the $SU(4)$
theory} together with $U(1),SU(2),SU(3)$ parts of the Standard Model at the common scale $\Lambda_{GUT}\sim 10^{14}\, GeV.$

Qualitatively, this idea is attractive because it naturally explains why the scale of this theory must be
much higher than the $SU(3)$ scale, $\Lambda_{QCD}$ a fraction of $GeV$. Indeed, with larger $N_c$
its coupling runs more quickly, and thus it gets strong closer to $\Lambda_{GUT}$.

More quantitatively, using the well known one-loop beta function coefficient 
$ b={11\over 3}N_c -{2 \over 3}N_f$
one has for QCD with $N_f=6$ (we ignore part of scales in which the effective $N_f$ is smaller than that)
$ b_{QCD}=7$, while for the theory in question it turns out to be exactly twice that, $b(N_c=4,N_f=2)=14$.  
This will put the scale of this theory exactly in between  $\Lambda_{QCD}$ and $\Lambda_{GUT}$
$$ \Lambda(N_c=4,N_f=2)=\sqrt{\Lambda_{QCD}\Lambda_{GUT} }\sim 10^7\, GeV$$

Of course, there can be much larger number of flavors, which would reduce this scale. It is also
possible that there is no Grand Unification. Experimental limits on this scale -- basically from the non-observed at LHC $SU(4)$ pions -- are in hundreds of GeV.

Anyway, if the scale  is high, neutral baryons have extremely small cross section \cite{Appelquist:2014jch}, even if the quarks themselves have
nonzero electroweak charges, similar to those of the usual quarks.

\section{The baryon clustering and the bosonic nucleosynthesis}  
Another qualitative idea which came  to my mind stems from basic nuclear physics as well as from the lessons from  Big Bang
Nucleorsyntesis (BBN). 

In QCD baryons can form nuclei, but not a single bound states of any number of neutrons exists. The physics of nuclei has a scale completely distinct from $\Lambda_{QCD}$:
a typical binding per nucleon is a factor 100 smaller than the nucleon mass, $B\sim M_N/100\sim 10\, MeV$.
This happens because of two fine tuned cancellations: (i) between attractive and repulsive forces, related to the specific
masses and couplings of scalar and vector mesons of the theory; and (ii) between the  potential and kinetic energies. 

Generically, there are no reasons for strong cancellations in the $SU(4)$ theory.
The second cancellation in particular,  due to Fermi energy induced by the fermionic nature of the SU(3) baryons, is no longer there.
The $SU(4)$ neutron (to be perhaps called $neutrone$, large neutron, opposite to neutrino, a small one), which one assumes to be the dark matter, are however $bosons$, and there is no Fermi energy. 
The $SU(4)$ nuclei may therefore be completely neutral, with
  cold Bose condensed
constituents, similar to systems of cold atoms at ultra-cold conditions.  
  
Let me remind that even very small binding  and very small baryon density at the BBN
time $n_N/n_\gamma \sim 10^{-10}$ do not stop nucleon clustering, for the following reason. Consider
for example the first reaction
$$ p+n\leftrightarrow d+\gamma$$
Huge excess of the photons drive the reaction to the left, unless it is stopped by 
the Boltzmann factor $exp(B_d/T)$. Yes, the deuteron binding is tiny $B_d=2.2 \, MeV$,
and yet, the temperature at the BBN time can be even much smaller! Slowly but surely,
the temperature of the Universe decreases, and when  the Boltzmann factor
gets large enough  $$exp(B_d/T)>10^{10} $$
the reaction proceeds to the right, saving large fraction of the neutrons by putting them inside
the deuterons (and eventually other nuclei).   We  are confident that this mechanism works well, as it produces
the deuteron fraction observed today. Furthermore, we can deduce from it rather strong limitations
on the variation of fundamental constants between now and the BBN time \cite{Flambaum:2002wq}.

Similarly, cosmologically produced SU(4) ``neutrones"  at the deconfinement transition of that theory will have 
long time to cluster, during eras with subsequently decreasing temperature,
eventually into larger and larger neutral ``drops".
Note further, that
unlike in the QCD case, there is no global Coulomb energy to limit the sizes of these clusters,
so it will be just limited by the available time and the rate of clustering. 

Incidentally, the author is now involved in studies of baryonic clustering at the freezeout stages of heavy ion collisions
\cite{clustering}.
The corresponding temperature is $T_f \sim 100 \, MeV$, high enough to melt any nuclei. And yet, interbaryon forces
do induce  clustering, which we study with classical molecular dynamics and which lead to 
correlations detected in the baryon number distributions, in spite of the fact that the time available is 
very short, $\sim 10 fm/c$, as compared to cosmological ones available for BBN. 
Clustering rate can have vastly different scales and be partially effective 
 even at a very long timescales. For example, globular clusters of stars in galaxies
 continue to capture stars for billions of years, becoming rather large, and yet still
  capturing only a small fraction of stars.

In summary, if the the dark matter is made of neutral $SU(4)$ baryons,
their mass scale can be rather large. It will in this case
 most likely come in form of ``nuclear drops", maybe even macroscopically large ones.
 If so, there may be hopes to detect their interaction. 
 As a first practical step to study this hypothetical scenario, one needs to know
 the value of the deconfinement transition of the SU(4) theory in question, 
 and approximate forces between baryons, which can be calculated on the lattice
 today.\\\\

{\bf Acknowledgement}: The author thanks E.Rinaldi 
 for  a seminar which introduced him to the subject.

\end{document}